\documentclass[universe,article,accept,moreauthors,pdftex]{Definitions/mdpi}

\firstpage{1} 
\pubvolume{1}
\issuenum{1}
\articlenumber{0}
\pubyear{2020}
\copyrightyear{2020}
\history{}

\usepackage{dcolumn} 
\newcolumntype{.}{D{.}{.}{4}}
\usepackage{siunitx}
\usepackage{textcomp}
\usepackage{pgf-pie}
\usepackage{bm}

\Title{Benefit of New High-Precision LLR Data for the Determination of Relativistic Parameters}

\Author{Liliane Biskupek $^{1}$\orcidA{}, J\"urgen M\"uller $^{1}$\orcidB and Jean-Marie Torre $^{2}$\orcidC{}}

\AuthorNames{Liliane Biskupek, J\"urgen M\"uller and Jean-Marie Torre}

\address{
$^{1}$ \quad Institute of Geodesy, Leibniz University Hannover, Schneiderberg 50, 30167 Hannover, Germany\\
$^{2}$ \quad Université Côte d'Azur, Observatoire de la Côte d'Azur, CNRS, IRD, Géoazur, Caussols, France
}

\corres{Correspondence: biskupek@ife.uni-hannover.de}

\abstract{Since 1969, Lunar Laser Ranging (LLR) data have been collected by various observatories and analysed by different analysis groups. In the recent years, observations with bigger telescopes (APOLLO) and at infra-red wavelength (OCA) are carried out, resulting in a better distribution of precise LLR data over the lunar orbit and the observed retro-reflectors on the Moon. This is a great advantage for various investigations in the LLR analysis. The aim of this study is to evaluate the benefit of the new LLR data for the determination of relativistic parameters. Here we show current results for relativistic parameters like a possible temporal variation of the gravitational constant $\dot{G}/G_0 = (-5.0 \pm 9.6) \times 10^{-15} \, \mathrm{yr}^{-1}$, the equivalence principle with $\Delta\left(m_g/m_i\right)_{\mathrm{EM}} = (-2.1 \pm 2.4)\times10^{-14}$ and the PPN parameters $\beta-1 = (6.2 \pm 7.2) \times 10^{-5}$ and $\gamma-1 = (1.7 \pm 1.6) \times 10^{-4}$. The results show a significant improvement in the accuracy of the various parameters, mainly due to better coverage of the lunar orbit, better distribution of measurements over the lunar retro-reflectors, and last but not least, higher accuracy of the data. Within the estimated accuracies, no violation of Einstein's theory is found and the results set improved limits for the different effects.}

\keyword{lunar laser ranging, gravitational constant, equivalence principle, PPN parameters}

\begin{document}


\section{Introduction}
 
It was July 20th, 1969 when the astronauts of the Apollo 11 crew landed in the southern part of Mare Tranquillitatis on the Moon. They deployed the Apollo Lunar Surface Experiments Package, where the retro-reﬂector for Lunar Laser Ranging (LLR) is now the last operating part of the experiment. Until 1973, four further reflectors were deployed on the lunar surface: two reflectors by the astronauts of the Apollo 14 and 15 missions, and two reflectors mounted on the unmanned Soviet Lunokhod rovers. For more than 50 years there is continuous measuring of the distance between observatories on the Earth and retro-reflectors on the Moon. The measurement of round trip travel times between Earth and Moon with short laser pulses is challenging. The average number of returning photons is less than one per laser pulse \cite{Chabe2020,Murphy2013}, mainly because of the beam divergence of the laser pulses due to the atmospheric turbulence and diffraction effects of the reflectors \cite{Murphy2010}. Further signal loss occurs in the paths of the transmitting and detection optics, in the atmosphere and due to the reflectivity of the retro-reflectors \cite{Mueller2019}. A series of single measurements over 5-15 minutes is used to calculate a so-called normal point (NP) which is the observable in the LLR analysis \cite{Michelsen2010}. The observatories on the Earth, that were or are capable to range to the Moon are the Observatoire de la Côte d’Azur (OCA) in France, the McDonald observatory (MLRS) and the Apache Point Lunar Laser-ranging Operation (APOLLO) in the USA, the Lunar Ranging Experiment (LURE) of the Haleakala observatory on Hawaii, the Matera Laser Ranging Observatory (MLRO) in Italy and the Wettzell Laser Ranging System (WLRS) in Germany. From the end of the eighties OCA started to investigate the measurement of laser round trip travel time from the observatory to the Moon and back with laser emitting in the infra-red (IR) at a wavelength $\lambda = \SI{1,064}{nm}$, but the precision level of IR detection was insufficient \cite{Courde2017}. In 2015 that problem was solved and OCA is now able to use IR and green laser for the regular measurements. After renewals and improvements WLRS is also able to measure the Earth-Moon distance with IR wavelength from 2018 on \cite{Schreiber2019}. With the new technique measurements close to the new Moon and full Moon are possible. Furthermore ranging to the retro-reflectors at lower elevation angles is possible and a better distribution of the measurements over the reflectors is achieved \cite{Chabe2020}. All that results in an improved coverage of the lunar orbit and is a big benefit in the analysis of the data and for the determination of various parameters. First tests of the universality of free-fall with IR data are promising \cite{Viswanathan2018}.

In Germany, from the early 80ies on, the software package LUNAR (LUNar laser ranging Analysis softwaRe) has been developed to study the Earth-Moon system and to determine several related model parameters \cite{Egger1985,Gleixner1986,Bauer1989, Muller1991}. Research covered physical libration and orbit of the Moon, coordinates of observatories and retro-reflectors, Earth orientation parameters. With special modifications of the model and software it is possible to investigate Einstein's theory of relativity by the determination of various parameters. Here, hard constraints for a possible violation of Einstein's theory are given by LLR results, like the equivalence principle, variation of the gravitational constant G, and the parametrized post-Newtonian (PPN) parameters $\beta$ and $\gamma$ \cite{Hofmann2018a,Zhang2020,Viswanathan2018}. By including the new high-precision NP measured with IR wavelength into the LLR analysis, improvements for various parameters are expected. The benefit of that NP is investigated here in more detail for the parameters related to testing Einstein's theory.

\section{LLR analysis}

The analysis model used in LUNAR is based on Einstein’s theory of relativity. It is fully relativistic and complete up to the first post-Newtonian ($1/c^2$) level. To take advantage of the high-precession NP that can be obtained with an accuracy of several millimetres \cite{Murphy2013}, the LUNAR software was updated continuously \cite{Biskupek2015,Hofmann2017}. Now the LLR analysis model take into account, among other things, gravitational effects of the Sun and planets with the Moon as extended body, the higher-order gravitational interaction between Earth and Moon as well as effects of the solid Earth tides on the lunar motion. The basis for the modelled lunar rotation is a 2-layer core-mantle model according to the DE430 ephemeris. A recent overview of LUNAR is given in \cite{Hofmann2018a}, a detailed description can be found in \cite{Muller2014a}.

The measured laser travel time $\tau_{meas}$ from Earth to Moon and back gives, together with the speed of light $c$, the length of the path (forth and back)

\begin{equation}
	\rho_{meas} = \tau_{meas} \cdot c \,. 
	\label{eq:rho_meas}
\end{equation}

\begin{figure}[H]
	\centering
	\def\svgwidth{0.5\linewidth}
	\input{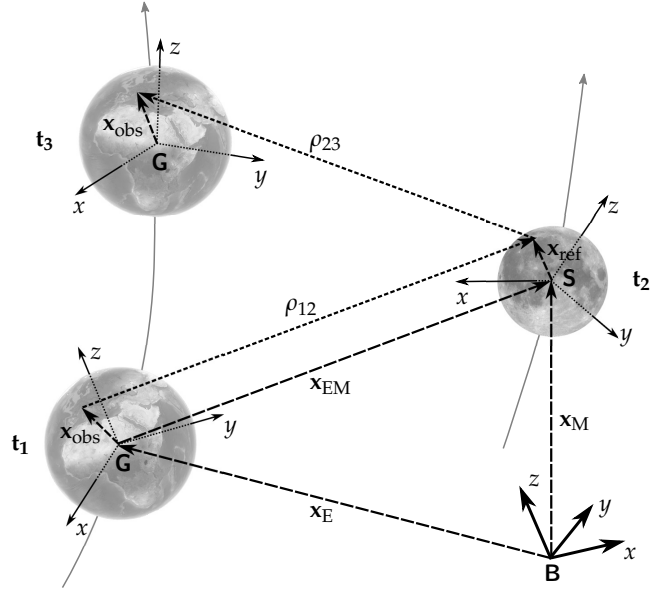}
	\caption{Scheme of a LLR measurement in the moving Earth-Moon system with the solar system barycenter B, geocenter G and selenocenter S, and the transmitting time $t_1$, reflection time $t_2$ and detection time $t_3$. The vectors are $\mathbf{x}_{\mathrm{E}}$ as barycentric vector to the geocenter at time $t_1$, $\mathbf{x}_{\mathrm{M}}$ as barycentric vector to the selenocenter at time $t_2$, $\mathbf{x}_{\mathrm{EM}} = \mathbf{x}_{\mathrm{M}}-\mathbf{x}_{\mathrm{E}}$ as Earth-Moon vector of the outgoing light path, $\mathbf{x}_{\mathrm{obs}}$ as vector geocenter-observatory in the barycentric system at the times $t_1$ and $t_3$, $\mathbf{x}_{\mathrm{ref}}$ as vector selenocenter-reflector in the barycentric system at the time $t_2$. $\rho_{12}$, $\rho_{23}$ as length of outgoing and incoming light path. $x,y,z$ indicate the different reference systems of the bodies and the solar system \cite{Hofmann2018a}.}
	\label{fig:measconfig}
\end{figure}

Figure \ref{fig:measconfig} shows the principle of LLR measurements in the moving Earth-Moon system. In a weighted least-squares adjustment the measured length of the light path is compared with the modelled length computed as

\begin{equation}
	\rho_{\mathrm{model}} = \rho_{12}+\rho_{23}+\Delta\tau \cdot c
	\label{eq:rho_model}
\end{equation}

where $\rho_{12}$ denotes the light path between the observatory at time $t_1$ and the reflector at time $t_2$ with

\begin{equation}
  \rho_{12}=\lvert\mathbf{x}_{\mathrm{EM}}(t_{1,2})+\mathbf{x}_{\mathrm{ref}}(t_2)-\mathbf{x}_{\mathrm{obs}}(t_1)\rvert
	\label{eq:rho12}
\end{equation}

and the light path $\rho_{23}$ between the reflector at time $t_2$ and the observatory at time $t_3$ with

\begin{equation}
	\rho_{23}=\lvert\mathbf{x}_{\mathrm{ME}}(t_{2,3})-\mathbf{x}_{\mathrm{ref}}(t_2)+\mathbf{x}_{\mathrm{obs}}(t_3)\rvert\,.
\label{eq:rho23}
\end{equation}

$\Delta\tau$ of equation \eqref{eq:rho_model} takes into account corrections of the light travel time, like the delay due to the light propagation through the gravitational potential of Sun and Earth \cite{Shapiro1964,Moyer1971}, an atmospheric delay \cite{Mendes2002,Mendes2004}, a synodic modulation of the lunar orbit due to radiation pressure \cite{Vokrouhlicky1997} and some time- and station-dependent biases.

For the calculation of equations \eqref{eq:rho12} and \eqref{eq:rho23} the positions of the observatories and retro-reflectors are needed in an inertial reference system, here the barycentric celestial reference system (BCRS) with the barycentric dynamical time (TDB) is selected. Therefore in a first step, effects at the coordinates of the observatories and retro-reflectors will be considered according to \cite{Petit2010}, e.g. tidal effects due to the atmosphere, ocean and solid Earth, and the tectonic movement in the respective body-fixed reference system. In a second step, the transformations from the body-fixed systems to the BCRS are carried out. Here the needed rotation of the Earth is modelled as defined in \cite{Biskupek2015,Petit2010}, the rotation of the Moon is computed simultaneously with the translation corresponding to \cite{Folkner2014}. The barycentric position and velocity vectors of Earth and Moon are derived from an ephemeris computation of the solar system bodies (all planets, the Moon and the largest asteroids). Initial values for the computation are taken from the DE421 ephemeris \cite{Folkner2009a}.

\begin{figure}[H]
	\def\svgwidth{\linewidth}
	\input{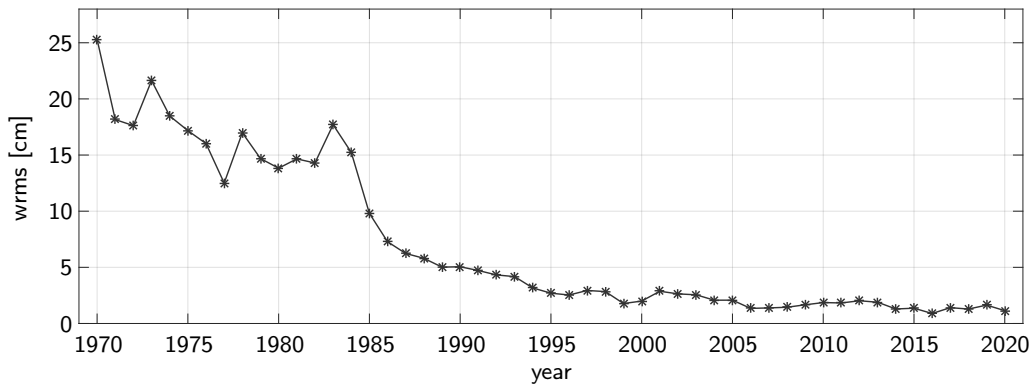}
	\caption{Annual weighted rms (wrms) of the one-way post-fit residuals for \num{27485} NP for the time span April 1970 to April 2020.}
	\label{fig:wrms}
\end{figure}	

The measured NP serve as observations in the analysis. They are treated as uncorrelated for the stochastic model of the least squares adjustment and are weighted according to their accuracy. The adjustment is done in a Gauss-Markov model where up to 250 unknown parameters can be determined with their uncertainties. As result of the analysis one gets the post-fit residuals, of which the weighted root-mean-square (wrms) is given in Figure \ref{fig:wrms} for each year. Beginning in 1970 with a wrms of more than \SI{25}{cm}, improvements in the measurement system lead to a decreasing wrms. Since 2006 it is about \SI{2}{cm} or less.

\section{Distribution of the normal points}

The distribution of LLR NP has a big impact on the determination of various parameters. Furthermore, non-uniform data distribution are one reason for correlations between solution parameters \cite{Williams2009b}. Therefore, the distribution of the LLR data is investigated in more detail below with respect to each of the observatories, retro-reflectors, and synodic angle. For the current study, NP for the period April 1970 to April 2020 were used. In a pre-analysis, all were investigated for possible outliers. Outliers are defined as NP whose residuals for the Earth-Moon distance exceed a limit of a few decimetres. Where the limit lies is determined differently for each observatory because they observe with different accuracies. Sometimes NP of one or more nights are shifted by the same amount, e.g. due to calibration problems during the measurement. Then a correction value is introduced into the analysis for this period. Furthermore, the standardized normal distribution is used for the evaluation of the outliers. If this distribution exceeds a certain value, the NP is also classified as an outlier and not included in the further analysis. After the pre-analysis, 17 NP of the current data set were identified as outliers. Thus, \num{27485} NP for the period April 1970 to April 2020 are included in the investigation, \num{22021} measured with green and \num{5464} with IR laser light. 

\begin{figure}[H]
  \def\svgwidth{\linewidth}
  \input{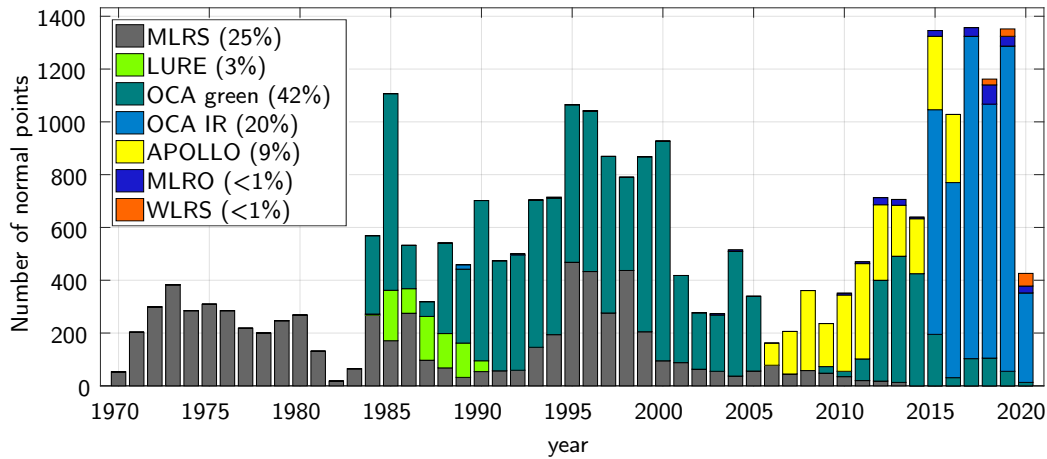}
	\caption{Distribution of the \num{27485} normal points over the the time span April 1970 - April 2020. In the legend the percentages of the contribution of the respective observatories are given. The three observatories McDonald, MLRS1 and MLRS2 are linked in the analysis and listed here as MLRS.}
	\label{fig:distributionNP}
\end{figure}

\begin{figure}[H]
\centering
\small\sffamily
	\begin{tikzpicture}
		\pie[rotate = -10, radius = 2, color = {gray!90, blue!50!green, yellow!80, green!30!blue, orange!90!red}, text = legend]
    {12/Apollo 11,  11/Apollo 14, 67/Apollo 15, 6/Lunokhod 2, 4/Lunokhod 1}
	\end{tikzpicture}
	\begin{tikzpicture}
		\pie[rotate = -10, radius = 2, color = {gray!90, blue!50!green, yellow!80, green!30!blue, orange!90!red}]
    {18/,  16/, 33/, 16/, 17/}
	\end{tikzpicture}
	\caption{Distribution of all NP as measurement to the respective reflectors for the whole data span of LLR (left) and for the year 2019 (right).}
	\label{fig:NPref}
\end{figure}
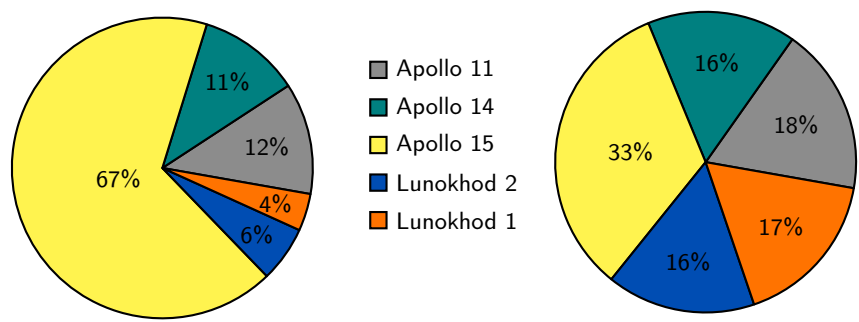

Figure \ref{fig:distributionNP} shows the temporal distribution of the measured NP over the last 50 years. One can see in the legend, that more then \SI{60}{\percent} of the NP were observed by OCA (\SI{42}{\percent} with green and  \SI{20}{\percent} with IR laser light). In the last years only OCA and APOLLO provided regular NP, some NP also came from MLRO and WLRS. For the year 2019, \SI{91}{\percent} of the NP were measured by OCA in IR. It is clear, that, with this distribution of NP, the analysis is dominated by the OCA NP. Looking at the distribution of the NP according to the respective reflectors, Figure \ref{fig:NPref} shows it for the whole time span in the left pie. Here the clear domination of Apollo 15 is obvious. Because Apollo 15 gives the strongest reflected signal due to its large size, it was more often tracked by the observatories in the past. This was not beneficial for data analysis. In recent years and with IR NP the situation has improved considerably. For 2019 (shown in Figure \ref{fig:NPref}, right pie) all retro-reflectors were measured approximately equally often, because of the advantage of IR laser light \cite{Chabe2020}. That is also a big advantage for the analysis, especially for the determination of the lunar libration. 

\begin{figure}[H]
  \def\svgwidth{\linewidth}
  \input{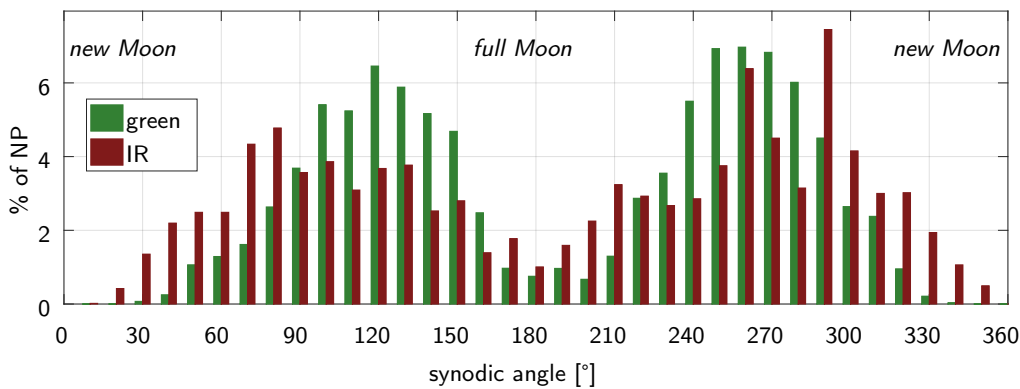}
		\caption{Distrubution of the NP over the synodic month. Given are the percentage of NP of the total number of measurements with the specific laser color. Full and new Moon indicates the phases of the Moon.}
	\label{fig:barSyn}
\end{figure}

With the better performance of the measurements now also ranging near new and full Moon is possible \cite{Chabe2020} for OCA and WLRS. This leads to a better coverage of the lunar orbit over the synodic month. The synodic month is the time span, when Sun, Earth, and Moon are in the same constellation again. To illustrate the better coverage, Figure \ref{fig:barSyn} shows the percentage of the NP measured for a specific angle of the synodic month. In green the measurements with green laser light and in red measurements with IR laser light are given. In the past with only green NP, there were gaps in the phases of new and full Moon. Now there are many more observations in IR and the advantage of it is obvious. The more equal distribution of the NP over the synodic month leads to a better coverage of the lunar orbit and is a big benefit for the determination of various parameters.

\section{Relativistic parameters}

The analysis of LLR observations is based on Einstein's theory of relativity. Thus, the Einstein-Infeld-Hoffmann equation, the signal propagation in the gravitational field of Earth and Sun, the temporal and spatial reference systems as well as their respective transformations are formulated relativistically \cite{Muller2014a}. By modification of the Einstein-Infeld-Hoffmann equation the relativistic parts are examined in more detail. Recent results, e.g. on the equivalence principle, Yukawa term, metric parameters and geodetic precession, can be found in publications \cite{Zhang2020,Hofmann2018a,Viswanathan2018,Williams2012}. \cite{Bourgoin2016} used LLR data to test parameters of the standard-model extension (SME). The various relativistic model contributions cause significant periodic variations, e.g. annual, monthly, linked to the node of the moon and combined periods in the Earth-Moon distance, through which it is possible to distinguish them from each other \cite{Muller2008}. Due to the large distance between Earth and Moon and the effect of the bodies in the solar system, the relativistic effects in the measured Earth-Moon distance are larger than, e.g. in distance measurements to satellites (SLR) \cite{Muller2008a}. This is a great advantage of LLR. Also, the long time span of LLR data (> \num{50} years) is very beneficial to determine and de-correlate certain relativistic parameters.

To determine relativistic parameters in the LLR analysis a two step strategy is used. In the first step the non-relativistic Newtonian parameters of the LUNAR model are calculated in a so called standard solution. Here the relativistic parameters are fixed to their values of Einstein's theory. The second step allow the estimation of individual relativistic parameters together with the Newtonian ones.   

In the following subsections different relativistic effects are investigated, e.g. the equivalence principle, the temporal variation of the gravitational constant and the PPN parameters $\gamma$ and $\beta$. The aim is to find out to what extent the higher precision IR NP have an impact on the estimation of the related parameters compared to the results of \cite{Hofmann2018a}, where a shorter time span (April 1070 - January 2015) with \num{20856} NP and much less IR NP were used. Table \ref{tab:summaryresults} gives an overview with the results of \cite{Hofmann2018a} and those of the current study. The basics of estimating the relativistic parameters are already given in \cite{Hofmann2018a} and are only briefly discussed here.

\begin{table}[H]
	\caption{Values for relativistic parameters from two different estimations. In the middle column results of \cite{Hofmann2018a}, in the right column results of the current estimation.}
	\begin{center}
		\begin{tabular}{ccc} \toprule 
			\multicolumn{1}{c}{parameter} & \multicolumn{1}{c}{\citet{Hofmann2018a}} & \multicolumn{1}{c}{current analysis}   \\ \midrule
			$\Delta \frac{m_g}{m_i}$ & $(-3.0 \pm 5.0) \times 10^{-14}$ & $(-2.1 \pm 2.4) \times 10^{-14}$   \\[0.1cm] 
			$\dot{G}/G$ &  $(7.1 \pm 7.6) \times 10^{-14} \ y^{-1}$ & $(-5.0 \pm 9.6) \times 10^{-15} \ y^{-1}$   \\ [0.1cm]
			$\ddot{G}/G$ & $(1.6 \pm 2.0) \times 10^{-15} \ y^{-2}$ & $(1.6 \pm 2.0) \times 10^{-16} \ y^{-2}$   \\ [0.1cm]
			$\gamma - 1$ &  $(-1.2 \pm 1.2) \times 10^{-4}$ & $(1.7 \pm 1.6) \times 10^{-4}$   \\ [0.1cm]
			$\beta - 1$ &  $(-8.7 \pm 9.0) \times 10^{-5}$ & $(6.2 \pm 7.2) \times 10^{-5}$   \\ \bottomrule 
		\end{tabular}
  \end{center}
	\label{tab:summaryresults}
\end{table}

\subsection{Equivalence principle}

The equivalence principle (EP) dates back to the 17th century when Galileo Galilei studied the acceleration of two bodies in free fall and found that in the same gravitational field it is independent of their shape, mass and composition \cite{Roll1964}. The second axiom of Isaac Newton states that the force $F$ results from the multiplication of an acceleration $a$ and the inertial mass $m_i$ as $F = m_i * a$. In the gravitational field of the Earth Newton's law of gravitation is $F = m_g * g$. That leads to the equivalence of the inertial mass $m_i$ and the gravitational mass $m_g$. If the equivalence principle is valid, the ratio $m_i/m_g$, which is called the weak equivalence principle (WEP), is equal to 1. The comparison of the free-fall accelerations of two bodies ($a_1, a_2$) leads to the test of the equivalence principle as

\begin{equation}
	\frac{\Delta a}{a}=\frac{2(a_1-a_2)}{a_1+a_2}= \frac{(m_g/m_i)_1-(m_g/m_i)_2}{[(m_g/m_i)_1+(m_g/m_i)_2]/2}\approx
		\left(\frac{m_g}{m_i}\right)_1-\left(\frac{m_g}{m_i}\right)_2=\Delta\frac{m_g}{m_i}\,.
		\label{eq:wep}
\end{equation}

A violation would lead to a different acceleration of the bodies in the same gravitational field. To investigate the WEP on Earth, sensitive torsion balances and test bodies made of different compositions like beryllium and titan \cite{Schlamminger2008}, and rubidium and potassium \cite{Albers2020} are used. Recent results from the MICROSCOPE satellite mission confirmed the WEP at the level of $\Delta m_g/m_i = (4 \pm 12) \times 10^{-15}$ \cite{Touboul2019}. From the analysis of LLR data between 1969 and 2017 \cite{Viswanathan2018} estimated $\Delta m_g/m_i = (-3.8 \pm 7.1) \times 10^{-14}$. 

In Einstein's gravitational theory the WEP is extended to the strong equivalence principle (SEP) due to the gravitational self-energy $U$ of the bodies. For bodies with astronomical sizes, like Earth and Moon, the SEP can be tested \cite{Nordtvedt1968} and parametrised with the Nordtvedt parameter $\eta$ by

\begin{equation}
	\frac{m_g}{m_i} = 1 + \eta \frac{U}{Mc^2}
	\label{eq:nordtvedt}
\end{equation} 

with self energy $U$ and mass $M$ for the respective body and the speed of light $c$. In Einstein's theory it holds $\eta = 0$. By analysing the LLR data a combined test of the SEP and WEP is possible. Here Earth and Moon are test bodies in the gravitational field of the Sun with gravitational self-energies and different composition. A violation of the EP would cause an additional acceleration of the Moon into the direction of the Sun. 

For the investigation of a possible violation of the EP with LLR data there are, according to \cite{Hofmann2018a}, two different way which leads to similar results. Here the focus is on the determination via an additional acceleration of the Moon $\ddot{\mathbf{x}}_{\mathrm{mgmi}}$ with the relative coordinates between Sun and Moon $\mathbf{x}_{\mathrm{SM}}$ and the distance $r_{\mathrm{SM}}$ to the Sun, where the largest perturbation is given by 

\begin{equation}
	\ddot{\mathbf{x}}_{\mathrm{mgmi}}=\Delta\left(\frac{m_g}{m_i}\right)_{\mathrm{EM}}GM_{\mathrm{Sun}}\frac{\mathbf{x}_{\mathrm{SM}}}{r_{\mathrm{SM}}^3}\,.
	\label{eq:mgmi_mond}
\end{equation}

$GM_{\mathrm{Sun}}$ denotes the gravitational constant times the mass of the Sun. That method keeps the interaction with all other forces in the calculation of the ephemeris. 

The result of \cite{Hofmann2018a} for the EP test was

\begin{equation*}
	\Delta\left(\frac{m_g}{m_i}\right)_{\mathrm{EM}} = (-3.0 \pm 5.0)\times10^{-14}
\label{eq:mgmi_res}
\end{equation*} 
 
with correlations of up to \SI{60}{\percent} with $GM_{EM}$ because of the dependence on the synodic angle $D$ \cite{Mueller1998}. There are also correlations of up to \SI{60}{\percent} with the X-components of the reflector coordinates. Compared to the result of this study with

\begin{equation*}
	\Delta\left(\frac{m_g}{m_i}\right)_{\mathrm{EM}} = (-2.1 \pm 2.4)\times10^{-14}
\label{eq:mgmi_res1}
\end{equation*} 

the accuracy was improved and the correlations to $GM_{EM}$ and the reflector coordinates decreased to \SI{40}{\percent}. Here the better coverage of the LLR NP over the synodic angle, shown in Figure \ref{fig:barSyn}, is a clear benefit for the determination of the EP parameter.

\cite{Zhang2020} investigated a possible violation of the EP due to assumed dark matter in the galactic centre which would cause an Earth–Moon
range oscillation with a sidereal month period. The amplitude for such an oscillation, determined from LLR post-fit residuals, was found to be $A = 0.6 \pm 1.0 \ \si{\milli\metre}$. The investigation also shows that a good orbit coverage with high precision data is more relevant for the EP test than the overall number of LLR data or a long time span. This underlines the improved values and correlations for the EP test, which shows no violation of Einstein's theory within the given error bars.

\subsection{Temporal variation of the gravitational constant}

From Einstein's general theory of relativity, it follows that the gravitational constant $G$ is a temporally and spatially invariable quantity \cite{Einstein1916}. According to the investigations of \cite{Brans1961} and \cite{Peebles1962}, however, the existence of alternative theories is possible, which allow a variation of the gravitational constant. One of the best known is the Brans-Dicke theory, a scalar-tensor theory. It is an extension of Einstein's theory with additional scalar fields \cite{Brans1961}. Recent studies by \cite{Sanders2010} and \cite{Steinhardt2010} confirm the considerations that a variation of the gravitational constant in the range from $\dot{G}/G_0 = 10^{-11} \, \mathrm{yr}^{-1}$ to $\dot{G}/G_0 = -10^{-14} \, \mathrm{yr}^{-1}$ might be possible. According to the remarks of \cite{Will1972}, there are also theories which admit so-called preferred reference systems. Also in such systems, a time dependence of $G$ would be possible. The recent upper bounds for a non-zero value of $\dot{G}$ by using LLR data come from the analysis of the ephemeris of the solar system bodies with $\dot{G}/G_0 = 7 \times 10^{-14} \, \mathrm{yr}^{-1}$ \cite{Pitjeva2014} and $\dot{G}/G_0 = 2 \times 10^{-13} \, \mathrm{yr}^{-1}$ \cite{Fienga2015}. From the analysis of MESSENGER data, \cite{Genova2018} get an upper limit for $\dot{G}/G_0 < 4 \times 10^{-14} \, \mathrm{yr}^{-1}$.

The estimation of a linear and quadratic part of the gravitational constant as a function of time is done in the analysis of LLR data with 

\begin{equation}
	G(t) = G_0 \left(1 + \frac{\dot{G}}{G_0}\Delta t + \frac{1}{2}\frac{\ddot{G}}{G_0}\Delta t^2 \right)
	\label{eq:G}
\end{equation}

as part of the ephemeris calculation. In the standard solution $\dot{G} = \ddot{G} = 0$ is valid. The time difference $\Delta t$ results from the current calculation time and the beginning of the LLR measurements. The partial derivatives of $\dot{G}$ and $\ddot{G}$ needed for the adjustment in the Gauss-Markov model are calculated by numerical differentiation of the geocentric lunar ephemeris.  

The results of \cite{Hofmann2018a} for the temporal and quadratic variation were estimated as separate parameters in two calculations as  

\begin{align*}
	\frac{\dot{G}}{G_0} = & \ (7.1 \pm 7.6) \times 10^{-14} \, \mathrm{yr}^{-1} \,, \\ 
	\frac{\ddot{G}}{G_0} = & \ (1.6 \pm 2.0) \times 10^{-15} \, \mathrm{yr}^{-2} \,.
\end{align*}

For this results, the initial values of the lunar core rotation vector $\omega_c$ were fixed to their estimated standard solution values because of the high correlation of up to \SI{94}{\percent} with $\dot{G}$ and $\ddot{G}$. High correlations of up to \SI{83}{\percent} with some components of the station coordinates were reduced by introducing constraints on the estimated station coordinates. 

The determination of $\dot{G}$ and $\ddot{G}$ with more NP including a high number of IR NP from OCA resulted in 

\begin{align*}
	\frac{\dot{G}}{G_0} = & \ (-5.0 \pm 9.6) \times 10^{-15} \, \mathrm{yr}^{-1} \,, \\ 
	\frac{\ddot{G}}{G_0} = & \ (1.6 \pm 2.0) \times 10^{-16} \, \mathrm{yr}^{-2} \,
\end{align*}

for the separate calculation of the two parts of $G$. In the current calculation the correlations with parts of the core rotation vector $\omega_c$ were up to \SI{70}{\percent} and decreased compared to \cite{Hofmann2018a}. But they were high enough to affect the determination of $\dot{G}$ and $\ddot{G}$, therefore they were fixed to their estimated standard solution values. The correlations with station coordinates could be reduced to up to \SI{20}{\percent} compared to \cite{Hofmann2018a} and no constraints were used in the calculation. Also correlations with other parameters of the Earth-Moon system significantly decreased and are now at most \SI{40}{\percent} with the Z-component of the initial values of the lunar orbit. Here the benefit of the longer data span with accurate NP leads to the improvement and a better and independent determination of the linear and quadratic part of $G$.  

For the determination of the linear and quadratic part of $G$ together the values are

\begin{align*}
	\frac{\dot{G}}{G_0} = & \ (0.2 \pm 1.3) \times 10^{-14} \, \mathrm{yr}^{-1} \,, \\ 
	\frac{\ddot{G}}{G_0} = & \ (2.0 \pm 2.8) \times 10^{-16} \, \mathrm{yr}^{-2} \,.
\end{align*}

The parameters are correlated to each other with \SI{70}{\percent}. The correlations to other parameters were also higher than in the separate estimation. This leads to a less accurate determination of $\dot{G}$ and $\ddot{G}$. Nevertheless, the accuracies are in a similar range as for the separate estimation and the results underline the validity of Einstein's theory in the given limits.

\subsection{PPN parameters $\beta$ and $\gamma$}

In the framework of the parametrized post-Newtonian (PPN) approximation of Einstein's theory, the parameter $\beta$ indicates the non-linearity of
gravity and $\gamma$ the size of space-curvature \cite{Will1972}. Both values are equal to 1 in this theory. 

Recent analysis of MESSENGER data \cite{Genova2018} gives a value for $\beta -1 = (-1.6 \pm 1.8) \times 10^{-5}$. From the analysis of solar system ephemeris there are values at the level of $7 \times 10^{-5}$ for $\beta$ and $5 \times 10^{-5}$ for $\gamma$ \cite{Fienga2015}. \cite{Pitjeva2013} gets values of $\beta -1 = (-2 \pm 3) \times 10^{-5}$ and $\gamma -1 = (4 \pm 6) \times 10^{-5}$.	

From the analysis of LLR data \cite{Hofmann2018a} get values for the determination of $\beta$ and $\gamma$ via the Einstein-Infeld-Hoffmann equations of motion as

\begin{align*}
	\beta -1 = & (-8.7 \pm 9.0) \times 10^{-5}\,, \\
	\gamma -1 = & (-1.2 \pm 1.2) \times 10^{-4}\,.	
\end{align*}

Both values show high correlations of up to \SI{82}{\percent} to station coordinates and the Z-component of the lunar initial velocity. Further correlations are to additional rotations between the Earth-fixed and space-fixed reference systems of up to \SI{47}{\percent}. For this reasons, the additional rotation was fixed to the values of the standard solution and the station coordinates were constraint. 

In the current analysis the PPN parameters were determined to 

\begin{align*}
	\beta -1 = & (6.2 \pm 7.2) \times 10^{-5}\,, \\
	\gamma -1 = & (1.7 \pm 1.6) \times 10^{-4}\,.	
\end{align*}

The correlations to the station coordinates now are up to \SI{60}{\percent}, to the Z-component of the lunar initial velocity \SI{40}{\percent} and to the additional rotation \SI{30}{\percent}. All of them were reduced significantly. To make the results more comparable with those of \cite{Hofmann2018a}, the additional rotations were nevertheless fixed to the values of the standard solution. The remaining correlations are now up to \SI{40}{\percent} with the previously mentioned parameters. Compared to the results of \cite{Hofmann2018a} the values of $\beta$ improved slightly, $\gamma$ is on a similar level. The longer time span and IR NP are not as beneficial for the estimation of PPN parameters as for the other relativistic parameters shown, but the correlations decreased and there is still no violation of Einstein's theory.

\section{Summary and Outlook}

The aim of this study was the investigation of the benefit of high-precision IR LLR measurements for determining relativistic parameters in comparison to the results of \cite{Hofmann2018a}. The model of the Earth-Moon system remained the same between the two calculations of relativistic parameters. The only major changes come from the longer time span of the LLR data and from many more measurements in the IR. From the previous discussions, it is clear that the IR data provide a major advantage for the LLR analysis. The accuracies of the relativistic parameters could be improved due to the better coverage of the lunar orbit and the accuracy of the data itself. Another advantage is the decorrelation of the relativistic parameters with other parameters of the Earth-Moon system. A summary of the results can be found in Table \ref{tab:summaryresults}. So far, from the analysis of the LLR data, the assumptions of Einstein's theory of relativity have been confirmed, now with improved limits.

An expanded network of single corner-cube retro-reflectors (CCRs) to be placed on the lunar surface near the limbs and poles from the year 2022 on will improve the existing geometry. Such CCRs are also beneficial in terms of thermal resilience and increased return signal strength. This will improve the ranging accuracy by a factor of 10 to 100 \cite{Currie2013}. 

With the construction of the new LLR facility at Table Mountain Observatory (JPL’s Optical Communication Testbed Laboratory - OCTL) in California, for the first time, it will be possible to conduct differential LLR with an expected range precision of less than 30 micrometers - a factor of 200 better than the current accuracy \cite{Turyshev2018}. This opens new possibilities for improved analysis of the whole LLR parameter set. 

The improvements on the technical side and further measurements in IR will make it possible, for example, to investigate effects related to the deep lunar interior and rotation and to determine relativistic parameters with higher accuracy. Together with improved modelling of the lunar interior and rotation in the LUNAR software, this will significantly improve many parameters determined from the analysis of the LLR data.  

\vspace{6pt}

\funding{This research was funded by the Deutsche Forschungsgemeinschaft (DFG, German Research Foundation) under Germany’s Excellence Strategy – EXC-2123 QuantumFrontiers – 390837967.}

\acknowledgments{Current LLR data are collected, archived, and distributed under the auspices of the International Laser Ranging Service (ILRS) \cite{Pearlman2002}. We acknowledge with thanks that more than 50 years of processed LLR data has been obtained under the efforts of the personnel at the Observatoire de la Côte dAzur in France, the LURE Observatory in Maui, Hawaii, the McDonald Observatory in Texas, the Apache Point Observatory in New Mexico, the Matera Laser Ranging station in Italy and the Wettzell Laser Ranging Station in Germany.}

\reftitle{References}

\externalbibliography{yes}
\bibliography{LLR_arXiv}

\begin{thebibliography}{-------}
\providecommand{\natexlab}[1]{#1}

\bibitem[Chabé \em{et~al.}(2020)Chabé, Courde, Torre, Bouquillon, Bourgoin,
  Aimar, Albanèse, Chauvineau, Mariey, Martinot-Lagarde, Maurice, Phung,
  Samain, and Viot]{Chabe2020}
Chabé, J.; Courde, C.; Torre, J.M.; Bouquillon, S.; Bourgoin, A.; Aimar, M.;
  Albanèse, D.; Chauvineau, B.; Mariey, H.; Martinot-Lagarde, G.; Maurice, N.;
  Phung, D.H.; Samain, E.; Viot, H.
\newblock {Recent Progress in Lunar Laser Ranging at Grasse Laser Ranging
  Station}.
\newblock {\em Earth and Space Science} {\bf 2020}, {\em 7},~e2019EA000785.
\newblock
  doi:{\changeurlcolor{black}\href{https://doi.org/10.1029/2019EA000785}{\detokenize{10.1029/2019EA000785}}}.

\bibitem[Murphy(2013)]{Murphy2013}
Murphy, T.W.
\newblock {Lunar laser ranging: the millimeter challenge}.
\newblock {\em Reports on Progress in Physics} {\bf 2013}, {\em 76},~076901.

\bibitem[Murphy \em{et~al.}(2010)Murphy, Adelberger, Battat, Hoyle, McMillan,
  Michelsen, Samad, Stubbs, and Swanson]{Murphy2010}
Murphy, T.W.; Adelberger, E.G.; Battat, J.B.R.; Hoyle, C.D.; McMillan, R.J.;
  Michelsen, E.L.; Samad, R.L.; Stubbs, C.W.; Swanson, H.E.
\newblock {Long-term degradation of optical devices on the Moon}.
\newblock {\em Icarus} {\bf 2010}, {\em 208},~31--35.
\newblock
  doi:{\changeurlcolor{black}\href{https://doi.org/http://dx.doi.org/10.1016/j.icarus.2010.02.015}{\detokenize{http://dx.doi.org/10.1016/j.icarus.2010.02.015}}}.

\bibitem[M{\"u}ller \em{et~al.}(2019)M{\"u}ller, Murphy, Schreiber, Shelus,
  Torre, Williams, Boggs, Bouquillon, Bourgoin, and Hofmann]{Mueller2019}
M{\"u}ller, J.; Murphy, T.W.; Schreiber, U.; Shelus, P.J.; Torre, J.M.;
  Williams, J.G.; Boggs, D.H.; Bouquillon, S.; Bourgoin, A.; Hofmann, F.
\newblock {Lunar Laser Ranging: a tool for general relativity, lunar geophysics
  and Earth science}.
\newblock {\em Journal of Geodesy} {\bf 2019}, {\em 93},~2195--2210.
\newblock
  doi:{\changeurlcolor{black}\href{https://doi.org/10.1007/s00190-019-01296-0}{\detokenize{10.1007/s00190-019-01296-0}}}.

\bibitem[Michelsen(2010)]{Michelsen2010}
Michelsen, E.L.
\newblock {Normal point generation and first photon bias correction in APOLLO
  Lunar Laser Ranging}.
\newblock PhD thesis, University of California, San Diego,  2010.

\bibitem[Courde \em{et~al.}(2017)Courde, Torre, Samain, Martinot-Lagarde,
  Aimar, Albanese, Exertier, Fienga, Mariey, Metris, Viot, and
  Viswanathan]{Courde2017}
Courde, C.; Torre, J.M.; Samain, E.; Martinot-Lagarde, G.; Aimar, M.; Albanese,
  D.; Exertier, P.; Fienga, A.; Mariey, H.; Metris, G.; Viot, H.; Viswanathan,
  V.
\newblock {Lunar laser ranging in infrared at the Grasse laser station}.
\newblock {\em Astronomy and Astrophysics} {\bf 2017}, {\em 602},~A90.
\newblock
  doi:{\changeurlcolor{black}\href{https://doi.org/10.1051/0004-6361/201628590}{\detokenize{10.1051/0004-6361/201628590}}}.

\bibitem[Schreiber \em{et~al.}(2019)Schreiber, Eckl, Leidig, Bachem, Neidhart,
  and Sch{\"u}ler]{Schreiber2019}
Schreiber, K.U.; Eckl, J.J.; Leidig, A.; Bachem, T.; Neidhart, A.; Sch{\"u}ler,
  T.
\newblock {Lunar Laser Ranging: A small system approach}.
\newblock  AGU Fall Meeting Abstracts,  2019, Vol. 2019, pp. G31B--0647.

\bibitem[Viswanathan \em{et~al.}(2018)Viswanathan, Fienga, Minazzoli, Bernus,
  Laskar, and Gastineau]{Viswanathan2018}
Viswanathan, V.; Fienga, A.; Minazzoli, O.; Bernus, L.; Laskar, J.; Gastineau,
  M.
\newblock {The new lunar ephemeris INPOP17a and its application to fundamental
  physics}.
\newblock {\em Monthly Notices of the Royal Astronomical Society} {\bf 2018},
  {\em 476},~1877--1888.
\newblock
  doi:{\changeurlcolor{black}\href{https://doi.org/10.1093/mnras/sty096}{\detokenize{10.1093/mnras/sty096}}}.

\bibitem[Egger(1985)]{Egger1985}
Egger, D.
\newblock {Systemanalyse der Laserentfernungsmessung}.
\newblock PhD thesis, Technische Universität München,  1985.
\newblock {Deutsche Geod\"atische Kommission bei der Bayerischen Akademie der
  Wissenschaften, Reihe C, Nr. 311}.

\bibitem[Gleixner(1986)]{Gleixner1986}
Gleixner, H.
\newblock {Ein Beitrag zur Ephemeridenrechnung und Parametersch\"atzung im
  Erde-Mond-System}.
\newblock PhD thesis, Technische Universit\"at M\"unchen,  1986.
\newblock {Deutsche Geod\"atische Kommission bei der Bayerischen Akademie der
  Wissenschaften, Reihe C, Nr. 319}.

\bibitem[Bauer(1989)]{Bauer1989}
Bauer, R.
\newblock {Bestimmung von Parametern des Erde-Mond-Systems - Ein Beitrag zur
  Modellerweiterung und Bewertung, Ergebnisse -}.
\newblock PhD thesis, Technische Universit\"at M\"unchen,  1989.
\newblock {Deutsche Geod\"atische Kommission bei der Bayerischen Akademie der
  Wissenschaften, Reihe C, Nr. 353}.

\bibitem[M\"uller(1991)]{Muller1991}
M\"uller, J.
\newblock {Analyse von Lasermessungen zum Mond im Rahmen einer
  post-Newton'schen Theorie}.
\newblock PhD thesis, Technische Universit\"at M\"unchen,  1991.
\newblock {Deutsche Geod\"atische Kommission bei der Bayerischen Akademie der
  Wissenschaften, Reihe C, Nr. 383}.

\bibitem[Hofmann and Müller(2018)]{Hofmann2018a}
Hofmann, F.; Müller, J.
\newblock {Relativistic tests with lunar laser ranging}.
\newblock {\em Classical and Quantum Gravity} {\bf 2018}, {\em 35},~035015.
\newblock
  doi:{\changeurlcolor{black}\href{https://doi.org/10.1088/1361-6382/aa8f7a}{\detokenize{10.1088/1361-6382/aa8f7a}}}.

\bibitem[Zhang \em{et~al.}(2020)Zhang, M{\"{u}}ller, and Biskupek]{Zhang2020}
Zhang, M.; M{\"{u}}ller, J.; Biskupek, L.
\newblock {Test of the equivalence principle for galaxy's dark matter by lunar
  laser ranging}.
\newblock {\em Celestial Mechanics and Dynamical Astronomy} {\bf 2020}, {\em
  132},~25.
\newblock
  doi:{\changeurlcolor{black}\href{https://doi.org/10.1007/s10569-020-09964-6}{\detokenize{10.1007/s10569-020-09964-6}}}.

\bibitem[Biskupek(2015)]{Biskupek2015}
Biskupek, L.
\newblock {Bestimmung der Erdorientierung mit Lunar Laser Ranging}.
\newblock PhD thesis, Leibniz Universität Hannover,  2015.
\newblock {Deutsche Geodätische Kommission bei der Bayerischen Akademie der
  Wissenschaften, Reihe C, Nr. 742},
  doi:{\changeurlcolor{black}\href{https://doi.org/10.15488/4721}{\detokenize{10.15488/4721}}}.

\bibitem[Hofmann(2017)]{Hofmann2017}
Hofmann, F.
\newblock {Lunar Laser Ranging – verbesserte Modellierung der Monddynamik und
  Schätzung relativistischer Parameter}.
\newblock PhD thesis, Leibniz Universität Hannover,  2017.
\newblock Deutsche Geodätische Kommission bei der Bayerischen Akademie der
  Wissenschaften, Reihe C, Nr. 797.

\bibitem[M\"uller \em{et~al.}(2014)M\"uller, Biskupek, Hofmann, and
  Mai]{Muller2014a}
M\"uller, J.; Biskupek, L.; Hofmann, F.; Mai, E.
\newblock {Lunar laser ranging and relativity}. In {\em {Frontiers in
  relativistic celestial Mechanics}}; Kopeikin, S.M., Ed.; Walter de Gruyter:
  Berlin,  2014; Vol. 2: Applications and Experiments, pp. 103--156.

\bibitem[Shapiro(1964)]{Shapiro1964}
Shapiro, I.I.
\newblock {Fourth Test of General Relativity}.
\newblock {\em Physical Review Letters} {\bf 1964}, {\em 13},~789--791.
\newblock
  doi:{\changeurlcolor{black}\href{https://doi.org/10.1103/PhysRevLett.13.789}{\detokenize{10.1103/PhysRevLett.13.789}}}.

\bibitem[Moyer(1971)]{Moyer1971}
Moyer, T.D.
\newblock {Mathematical formulation of the Double Precision Orbit Determination
  Program DPODP}.
\newblock Technical Report JPL-TR-32-1527, Jet Propulsion Laboratory,
  California Institute of Technology, Pasadena, California,  1971.

\bibitem[Mendes \em{et~al.}(2002)Mendes, Prates, Pavlis, Pavlis, and
  Lanley]{Mendes2002}
Mendes, V.B.; Prates, G.; Pavlis, E.C.; Pavlis, D.E.; Lanley, R.B.
\newblock {Improved mapping functions for atmospheric refraction correction in
  SLR}.
\newblock {\em Geophysical Research Letters} {\bf 2002}, {\em 29},~1414.
\newblock
  doi:{\changeurlcolor{black}\href{https://doi.org/10.1029/2001GL014394}{\detokenize{10.1029/2001GL014394}}}.

\bibitem[Mendes and Pavlis(2004)]{Mendes2004}
Mendes, V.B.; Pavlis, E.C.
\newblock {High-accuracy zenith delay prediction at optical wavelengths}.
\newblock {\em Geophysical Research Letters} {\bf 2004}, {\em 31},~L14602.
\newblock
  doi:{\changeurlcolor{black}\href{https://doi.org/10.1029/2004GL020308}{\detokenize{10.1029/2004GL020308}}}.

\bibitem[Vokrouhlicky(1997)]{Vokrouhlicky1997}
Vokrouhlicky, D.
\newblock {A Note on the Solar Radiation Perturbations of Lunar Motion}.
\newblock {\em Icarus} {\bf 1997}, {\em 126},~293--300.
\newblock
  doi:{\changeurlcolor{black}\href{https://doi.org/10.1006/icar.1996.5652}{\detokenize{10.1006/icar.1996.5652}}}.

\bibitem[Petit and Luzum(2010)]{Petit2010}
Petit, G.; Luzum, B., Eds.
\newblock {\em {IERS Conventions 2010}}; Number~36 in IERS Technical Note,
  Verlag des Bundesamtes f\"ur Kartographie und Geod\"asie: Frankfurt am Main,
  2010.

\bibitem[Folkner \em{et~al.}(2014)Folkner, Williams, Boggs, Park, and
  Kuchynka]{Folkner2014}
Folkner, W.M.; Williams, J.G.; Boggs, D.H.; Park, R.S.; Kuchynka, P.
\newblock {The Planetary and Lunar Ephemerides DE430 and DE431}. In {\em The
  Interplanetary Network Progress Report}; Jet Propulsion Laboratory,
  California Institute of Technology: Passadena, Califonia,  2014; Vol. 42-196.

\bibitem[Folkner \em{et~al.}(2009)Folkner, Williams, and Boggs]{Folkner2009a}
Folkner, W.M.; Williams, J.G.; Boggs, D.H.
\newblock {The Planetary and Lunar Ephemeris DE 421}. In {\em The
  Interplanetary Network Progress Report}; Jet Propulsion Laboratory,
  California Institute of Technology: Passadena, Califonia,  2009; Vol. 42-178.

\bibitem[Williams \em{et~al.}(2009)Williams, Turyshev, and
  Boggs]{Williams2009b}
Williams, J.G.; Turyshev, S.G.; Boggs, D.H.
\newblock {Lunar Laser Ranging Tests of the Equivalence Principle with the
  Earth and Moon}.
\newblock {\em International Journal of Modern Physics D} {\bf 2009}, {\em
  18},~1129--1175.
\newblock
  doi:{\changeurlcolor{black}\href{https://doi.org/10.1142/S021827180901500X}{\detokenize{10.1142/S021827180901500X}}}.

\bibitem[Williams \em{et~al.}(2012)Williams, Turyshev, and Boggs]{Williams2012}
Williams, J.G.; Turyshev, S.G.; Boggs, D.H.
\newblock {Lunar laser ranging tests of the equivalence principle}.
\newblock {\em Classical and Quantum Gravity} {\bf 2012}, {\em 29},~184004.

\bibitem[Bourgoin \em{et~al.}(2016)Bourgoin, Hees, Bouquillon,
  Le~Poncin-Lafitte, Francou, and Angonin]{Bourgoin2016}
Bourgoin, A.; Hees, A.; Bouquillon, S.; Le~Poncin-Lafitte, C.; Francou, G.;
  Angonin, M.C.
\newblock {Testing Lorentz Symmetry with Lunar Laser Ranging}.
\newblock {\em Physical Review Letters} {\bf 2016}, {\em 117},~241301.
\newblock
  doi:{\changeurlcolor{black}\href{https://doi.org/10.1103/PhysRevLett.117.241301}{\detokenize{10.1103/PhysRevLett.117.241301}}}.

\bibitem[M\"uller \em{et~al.}(2008{\natexlab{a}})M\"uller, Williams, and
  Turyshev]{Muller2008}
M\"uller, J.; Williams, J.G.; Turyshev, S.G.
\newblock {Lunar Laser Ranging Contributions to Relativity and Geodesy}.
\newblock  {Lasers, Clocks and Drag-Free Control: Exploration of Relativistic
  Gravity in Space}; Dittus, H.; L\"ammerzahl, C.; Turyshev, S.G., Eds.;
  Springer: Berlin, Heidelberg,  2008; Vol. 349, {\em Astrophysics and Space
  Science Library}, pp. 457--472.
\newblock
  doi:{\changeurlcolor{black}\href{https://doi.org/10.1007/978-3-540-34377-6_21}{\detokenize{10.1007/978-3-540-34377-6_21}}}.

\bibitem[M\"uller \em{et~al.}(2008{\natexlab{b}})M\"uller, Soffel, and
  Klioner]{Muller2008a}
M\"uller, J.; Soffel, M.; Klioner, S.A.
\newblock {Geodesy and relativity}.
\newblock {\em Journal of Geodesy} {\bf 2008}, {\em 82},~133--145.
\newblock
  doi:{\changeurlcolor{black}\href{https://doi.org/10.1007/s00190-007-0168-7}{\detokenize{10.1007/s00190-007-0168-7}}}.

\bibitem[Roll \em{et~al.}(1964)Roll, Krotkov, and Dicke]{Roll1964}
Roll, P.G.; Krotkov, R.; Dicke, R.H.
\newblock {The equivalence of inertial and passive gravitational mass}.
\newblock {\em Annals of Physics} {\bf 1964}, {\em 26},~442--517.
\newblock
  doi:{\changeurlcolor{black}\href{https://doi.org/https://doi.org/10.1016/0003-4916(64)90259-3}{\detokenize{https://doi.org/10.1016/0003-4916(64)90259-3}}}.

\bibitem[Schlamminger \em{et~al.}(2008)Schlamminger, Choi, Wagner, Gundlach,
  and Adelberger]{Schlamminger2008}
Schlamminger, S.; Choi, K.Y.; Wagner, T.A.; Gundlach, J.H.; Adelberger, E.G.
\newblock {Test of the Equivalence Principle Using a Rotating Torsion Balance}.
\newblock {\em Physical Review Letters} {\bf 2008}, {\em 100},~041101.
\newblock
  doi:{\changeurlcolor{black}\href{https://doi.org/10.1103/PhysRevLett.100.041101}{\detokenize{10.1103/PhysRevLett.100.041101}}}.

\bibitem[Albers \em{et~al.}(2020)Albers, Herbst, Richardson, Heine, Nath,
  Hartwig, Schubert, Vogt, Woltmann, Lämmerzahl, Herrmann, Ertmer, Rasel, and
  Schlippert]{Albers2020}
Albers, H.; Herbst, A.; Richardson, L.L.; Heine, H.; Nath, D.; Hartwig, J.;
  Schubert, C.; Vogt, C.; Woltmann, M.; Lämmerzahl, C.; Herrmann, S.; Ertmer,
  W.; Rasel, E.M.; Schlippert, D.
\newblock {Quantum test of the Universality of Free Fall using rubidium and
  potassium}.
\newblock {\em The European Physical Journal D} {\bf 2020}, {\em 74},~145.
\newblock
  doi:{\changeurlcolor{black}\href{https://doi.org/10.1140/epjd/e2020-10132-6}{\detokenize{10.1140/epjd/e2020-10132-6}}}.

\bibitem[Touboul \em{et~al.}(2019)Touboul, M{\'{e}}tris, Rodrigues,
  Andr{\'{e}}, Baghi, Berg{\'{e}}, Boulanger, Bremer, Chhun, Christophe,
  Cipolla, Damour, Danto, Dittus, Fayet, Foulon, Guidotti, Hardy, Huynh,
  Lämmerzahl, Lebat, Liorzou, List, Panet, Pires, Pouilloux, Prieur, Reynaud,
  Rievers, Robert, Selig, Serron, Sumner, and Visser]{Touboul2019}
Touboul, P.; M{\'{e}}tris, G.; Rodrigues, M.; Andr{\'{e}}, Y.; Baghi, Q.;
  Berg{\'{e}}, J.; Boulanger, D.; Bremer, S.; Chhun, R.; Christophe, B.;
  Cipolla, V.; Damour, T.; Danto, P.; Dittus, H.; Fayet, P.; Foulon, B.;
  Guidotti, P.Y.; Hardy, E.; Huynh, P.A.; Lämmerzahl, C.; Lebat, V.; Liorzou,
  F.; List, M.; Panet, I.; Pires, S.; Pouilloux, B.; Prieur, P.; Reynaud, S.;
  Rievers, B.; Robert, A.; Selig, H.; Serron, L.; Sumner, T.; Visser, P.
\newblock {Space test of the equivalence principle: first results of the
  {MICROSCOPE} mission}.
\newblock {\em Classical and Quantum Gravity} {\bf 2019}, {\em 36},~225006.
\newblock
  doi:{\changeurlcolor{black}\href{https://doi.org/10.1088/1361-6382/ab4707}{\detokenize{10.1088/1361-6382/ab4707}}}.

\bibitem[Nordtvedt(1968)]{Nordtvedt1968}
Nordtvedt, K.
\newblock {Equivalence Principle for Massive Bodies. I. Phenomenology}.
\newblock {\em Physical Review D} {\bf 1968}, {\em 169},~1014--1016.
\newblock
  doi:{\changeurlcolor{black}\href{https://doi.org/10.1103/PhysRev.169.1014}{\detokenize{10.1103/PhysRev.169.1014}}}.

\bibitem[M\"uller and Nordtvedt(1998)]{Mueller1998}
M\"uller, J.; Nordtvedt, K.
\newblock {Lunar laser ranging and the equivalence principle signal}.
\newblock {\em Physical Review D} {\bf 1998}, {\em 58},~062001.
\newblock
  doi:{\changeurlcolor{black}\href{https://doi.org/10.1103/PhysRevD.58.062001}{\detokenize{10.1103/PhysRevD.58.062001}}}.

\bibitem[Einstein(1916)]{Einstein1916}
Einstein, A.
\newblock {Die Grundlage der allgemeinen Relativitätstheorie}.
\newblock {\em Annalen der Physik} {\bf 1916}, {\em 354},~769--822.
\newblock
  doi:{\changeurlcolor{black}\href{https://doi.org/10.1002/andp.19163540702}{\detokenize{10.1002/andp.19163540702}}}.

\bibitem[Brans and Dicke(1961)]{Brans1961}
Brans, C.; Dicke, R.H.
\newblock {Mach's Principle and a Relativistic Theory of Gravitation}.
\newblock {\em Physical Review} {\bf 1961}, {\em 124},~925--935.
\newblock
  doi:{\changeurlcolor{black}\href{https://doi.org/10.1103/PhysRev.124.925}{\detokenize{10.1103/PhysRev.124.925}}}.

\bibitem[Peebles and Dicke(1962)]{Peebles1962}
Peebles, P.J.; Dicke, R.H.
\newblock {Significance of Spatial Isotropy}.
\newblock {\em Physical Review} {\bf 1962}, {\em 127},~629--631.
\newblock
  doi:{\changeurlcolor{black}\href{https://doi.org/10.1103/PhysRev.127.629}{\detokenize{10.1103/PhysRev.127.629}}}.

\bibitem[Sanders \em{et~al.}(2010)Sanders, Gillies, and Schmutzer]{Sanders2010}
Sanders, A.; Gillies, G.; Schmutzer, E.
\newblock {Implications upon theory discrimination of an accurate measurement
  of the time rate of change of the gravitational “constant” G and other
  cosmological parameters}.
\newblock {\em Annalen der Physik} {\bf 2010}, {\em 522},~861--873.
\newblock
  doi:{\changeurlcolor{black}\href{https://doi.org/10.1002/andp.201010460}{\detokenize{10.1002/andp.201010460}}}.

\bibitem[Steinhardt and Wesley(2010)]{Steinhardt2010}
Steinhardt, P.J.; Wesley, D.
\newblock {Exploring extra dimensions through observational tests of dark
  energy and varying Newton's constant}.
\newblock arXiv:1003.2815v1,  2010.
\newblock Version from 14.03.2010.

\bibitem[Will and Nordtvedt(1972)]{Will1972}
Will, C.M.; Nordtvedt, K.
\newblock {Conservation Laws and Preferred Frames in Relativistic Gravity. I.
  Preferred-Frame Theories and an Extended PPN Formalism}.
\newblock {\em The Astrophysical Journal} {\bf 1972}, {\em 177},~757--774.
\newblock
  doi:{\changeurlcolor{black}\href{https://doi.org/10.1086/151754}{\detokenize{10.1086/151754}}}.

\bibitem[Pitjeva and Pitjev(2014)]{Pitjeva2014}
Pitjeva, E.V.; Pitjev, N.P.
\newblock {Development of planetary ephemerides EPM and their applications}.
\newblock {\em Celestial Mechanics and Dynamical Astronomy} {\bf 2014}, {\em
  119},~237--256.
\newblock
  doi:{\changeurlcolor{black}\href{https://doi.org/10.1007/s10569-014-9569-0}{\detokenize{10.1007/s10569-014-9569-0}}}.

\bibitem[Fienga \em{et~al.}(2015)Fienga, Laskar, Exertier, Manche, and
  Gastineau]{Fienga2015}
Fienga, A.; Laskar, J.; Exertier, P.; Manche, H.; Gastineau, M.
\newblock {Numerical estimation of the sensitivity of INPOP planetary
  ephemerides to general relativity parameters}.
\newblock {\em Celestial Mechanics and Dynamical Astronomy} {\bf 2015}, {\em
  123},~325--349.
\newblock
  doi:{\changeurlcolor{black}\href{https://doi.org/10.1007/s10569-015-9639-y}{\detokenize{10.1007/s10569-015-9639-y}}}.

\bibitem[Genova \em{et~al.}(2018)Genova, Mazarico, Goossens, Lemoine, Neumann,
  Smith, and Zuber]{Genova2018}
Genova, A.; Mazarico, E.; Goossens, S.; Lemoine, F.G.; Neumann, G.A.; Smith,
  D.E.; Zuber, M.T.
\newblock {Solar system expansion and strong equivalence principle as seen by
  the NASA MESSENGER mission}.
\newblock {\em Nature Communications} {\bf 2018}, {\em 9},~289.
\newblock
  doi:{\changeurlcolor{black}\href{https://doi.org/10.1038/s41467-017-02558-1}{\detokenize{10.1038/s41467-017-02558-1}}}.

\bibitem[Pitjeva and Pitjev(2013)]{Pitjeva2013}
Pitjeva, E.V.; Pitjev, N.P.
\newblock {Relativistic effects and dark matter in the Solar system from
  observations of planets and spacecraft}.
\newblock {\em Monthly Notices of the Royal Astronomical Society} {\bf 2013},
  {\em 432},~3431--3437.
\newblock
  doi:{\changeurlcolor{black}\href{https://doi.org/10.1093/mnras/stt695}{\detokenize{10.1093/mnras/stt695}}}.

\bibitem[Currie \em{et~al.}(2013)Currie, Dell'Agnello, {Delle Monache}, Behr,
  and Williams]{Currie2013}
Currie, D.G.; Dell'Agnello, S.; {Delle Monache}, G.O.; Behr, B.; Williams, J.G.
\newblock {A Lunar Laser Ranging Retroreflector Array for the 21st Century}.
\newblock {\em Nuclear Physics B} {\bf 2013}, {\em 243--244},~218--228.
\newblock
  doi:{\changeurlcolor{black}\href{https://doi.org/10.1016/j.nuclphysbps.2013.09.007}{\detokenize{10.1016/j.nuclphysbps.2013.09.007}}}.

\bibitem[Turyshev \em{et~al.}(2018)Turyshev, Shao, Hanh, Williams, and
  Trahan]{Turyshev2018}
Turyshev, S.G.; Shao, M.; Hanh, I.; Williams, J.G.; Trahan, R.
\newblock {Advanced Laser Ranging for high-precision science investigations}.
\newblock Talk,  2018.
\newblock 21st International Workshop on Laser Ranging, 5-9 November 2018,
  Canberra, Australia.

\bibitem[Pearlman \em{et~al.}(2002)Pearlman, Degnan, and
  Bosworth]{Pearlman2002}
Pearlman, M.R.; Degnan, J.J.; Bosworth, J.M.
\newblock {The International Laser Ranging Service}.
\newblock {\em Advances in Space Research} {\bf 2002}, {\em 30},~135--143.
\newblock
  doi:{\changeurlcolor{black}\href{https://doi.org/10.1016/S0273-1177(02)00277-6}{\detokenize{10.1016/S0273-1177(02)00277-6}}}.

\end{thebibliography}

\end{document}